\def\usecolorlinks{0}
\def\preprintmode{0}
\def\useitalicorbitals{0}
\def\useitalicspacegroups{1}
\def\buildbib{0}
\def\highlightchanges{0}

\if\preprintmode1
	\documentclass[aps,prb,preprint,%
		superscriptaddress,citeautoscript,showkeys,%
		endfloats*,floatfix,titlepage]%
		{revtex4-1}
\else
	\documentclass[aps,prb,reprint,%
		superscriptaddress,citeautoscript,showkeys,%
		floatfix,notitlepage]%
		{revtex4-1}
\fi

\usepackage{amsmath,amssymb,amsthm}         
\usepackage{textcomp,gensymb,mathtools}     
\usepackage{bm}                             

\usepackage{tabu}                           
\usepackage{multirow}                       

\usepackage{graphicx}
\graphicspath{ {./} }

\usepackage[pdftex,bookmarks,colorlinks,breaklinks]{hyperref}
\usepackage[usenames]{color}                
\definecolor{dullmagenta}{rgb}{0.4, 0, 0.4}
\definecolor{darkblue}{rgb}{0, 0, 0.4}
\if\usecolorlinks1
	\hypersetup{%
		linkcolor = red, 
		citecolor = blue, 
		filecolor = dullmagenta, 
		urlcolor = darkblue}
\else
	\hypersetup{%
		linkcolor = black, 
		citecolor = black,
		filecolor = black,
		urlcolor = black}
\fi

\if\highlightchanges1
	\definecolor{change}{rgb}{0.25, 0, 1}  
	\definecolor{CHANGE}{rgb}{0.25, 0, 1}  
	\newcommand{\change}[1]{\textcolor{change}{#1}}
\else
	\newcommand{\change}[1]{#1}            
	\definecolor{change}{rgb}{0, 0, 0}     
\fi

\makeatletter
\newcommand*{\IfBoldTF}{%
	\ifx\f@series\test@for@bold%
		\expandafter\@firstoftwo%
	\else%
		\expandafter\@secondoftwo%
	\fi%
}
\newcommand*{\test@for@bold}{bx}
\makeatother
\newcommand{\checkboldmath}[1]{\texorpdfstring%
	{\IfBoldTF{\ensuremath{\bm{#1}}}{\ensuremath{#1}}}%
	{\ensuremath{#1}}%
}

\newcommand{\SCO}{SrCoO\textsubscript{2.5}}
\newcommand{\HSCO}{HSrCoO\textsubscript{2.5}}
\newcommand{\Co}[1]{Co\textsubscript{(#1)}}
\newcommand{\Ox}[1]{O\textsubscript{(#1)}}

\newcommand{\MagE}{\checkboldmath{\Delta F_{\text{AFM}-\text{FM}}/N_{\text{Co}}%
	}}
\newcommand{\BohrM}{\checkboldmath{\mu_{\text{B}}}}
\newcommand{\Ang}{\r{A}}
\newcommand{\dgsign}{\ifmmode\text{\textdegree}\else\textdegree\fi}

\if\useitalicorbitals1
	\newcommand{\orb}[2]{\textit{#1}\textsubscript{\checkboldmath{#2}}}
\else
	\newcommand{\orb}[2]{\textrm{#1}\textsubscript{\checkboldmath{#2}}}
\fi
\if\useitalicspacegroups1
	\newcommand{\spacegrp}[1]{\textit{#1}}
\else
	\newcommand{\spacegrp}[1]{\textrm{#1}}
\fi
\newcommand{\ImaTwo}{\spacegrp{Ima2}}
\newcommand{\PmcTwoSubOne}{\spacegrp{Pmc2\textsubscript{1}}}
\newcommand{\PnaTwoSubOne}{\spacegrp{Pna2\textsubscript{1}}}

\begin{document}
\pagenumbering{arabic}

\title{Towards understanding the special stability of \SCO{} and \HSCO{}}
\date{\today}


\author{Sze-Chun \surname{Tsang}}
\author{Jingzhao \surname{Zhang}}
\author{Kinfai \surname{Tse}}
\author{Junyi \surname{Zhu}}
\email[Corresponding author; ]{jyzhu@phy.cuhk.edu.hk}
\affiliation{Department of Physics, the Chinese University of Hong Kong, %
	Shatin, New Territories, Hong Kong}

\hypersetup{%
	pdftitle = {%
		Towards understanding the special stability of \SCO{} and \HSCO{}},
	pdfauthor = {%
		Tsang, Sze-Chun; 
		Zhang, Jingzhao; 
		Tse, Kinfai; 
		Zhu, Junyi},
	pdfkeywords = {%
		strontium cobaltite; 
		hydrogen storage; 
		brownmillerite;
		electron-counting model; 
		DFT+U}
}


\begin{abstract}
Reversible hydrogen incorporation was recently attested 
[N. Lu, \emph{et al.}, Nature \textbf{546}, 124 (2017)] in \SCO{}, 
the brownmillerite phase (BM) of strontium cobalt oxide (SCO), 
opening new avenues in catalysis and energy applications. 
However, existing theoretical studies of {BM-SCO} are insufficient, 
and that of \HSCO{}, the newly-reported hydrogenated SCO ({H-SCO}) phase, 
is especially scarce. 
In this work, 
we demonstrate how the electron-counting model (ECM) can be used in %
	understanding the phases, 
particularly in explaining the stability of the oxygen-vacancy channels (OVCs), 
and in examining the Co valance problem. 
Using density-functional theoretical (DFT) methods, 
we analyze the crystalline, electronic, and magnetic structures of BM- and %
	{H-SCO}. 
Based on our structure search, 
we discovered stable phases with large bandgaps ({$>1$~eV}) for both {BM-SCO} %
	and {H-SCO}, 
agreeing better with experiments on the electronic structures. 
Our calculations also indicate limited charge transfer from H to O, 
which may explain the special stability of the {H-SCO} phase and the %
	reversibility of H incorporation observed in experiments. 
In contrary to the initial study, 
our calculations also suggest intrinsic antiferromagnetism (AFM) of {H-SCO}, 
showing how the measured ferromagnetism (FM) has possible roots in hole doping.
\end{abstract}

\maketitle


\section{Introduction}
\label{sect:Intro}



In 2017, reversible hydrogen incorporation and three-phase interconversion was %
	realized \cite{TriStateXform} in the brownmillerite (BM) phase of %
	strontium cobalt oxide (SCO, or strontium cobaltite), 
of a chemical formula \SCO{}, 
via ionic-liquid gating; 
and a new, hydrogenated phase of SCO ({H-SCO}, formula \HSCO{}) was discovered. 
Due to its exceptionally robust oxygen-vacancy channels (OVCs) which %
	facilitate ionic transport, incorporation, and extraction, 
{BM-SCO} has promising prospects in fuel cell technology, energy storage, and %
	catalysis. 
Such breakthrough in the study and manipulation of magnetic complex oxides is %
	also highly encouraging towards ongoing research in the field.

Illustration of the two phases are provided in the form of Fig.~%
	\ref{fig:StructIllust}, 
with the structures available in the \texttt{.cif} format in Supplemental \
	Material (SM) for the interested reader.%
	\footnote{See Supplemental Material at [URL] for additional %
		results on functional and $U$~value tests, 
		\change{phonon calculations on the {BM-SCO} phases,} 
		notes on the single-hydrogen {BM-SCO} configurations, 
		band-component analyses of the {BM-SCO} and {H-SCO} phases, 
		and \texttt{.cif} files of the structures.}%
To distinguish between the various cobalt (Co) and oxygen (O) sites, 
we named the Co centers according to their coordination numbers 
(\Co{6} is six-coordinated, etc.), 
and the O sites according to the metallic sites they are approximately %
	coplanar with 
[\Ox{6} for \Co{6}, \Ox{Sr} for strontium (Sr), etc.].
\begin{figure}
	\includegraphics[width=3.375in]{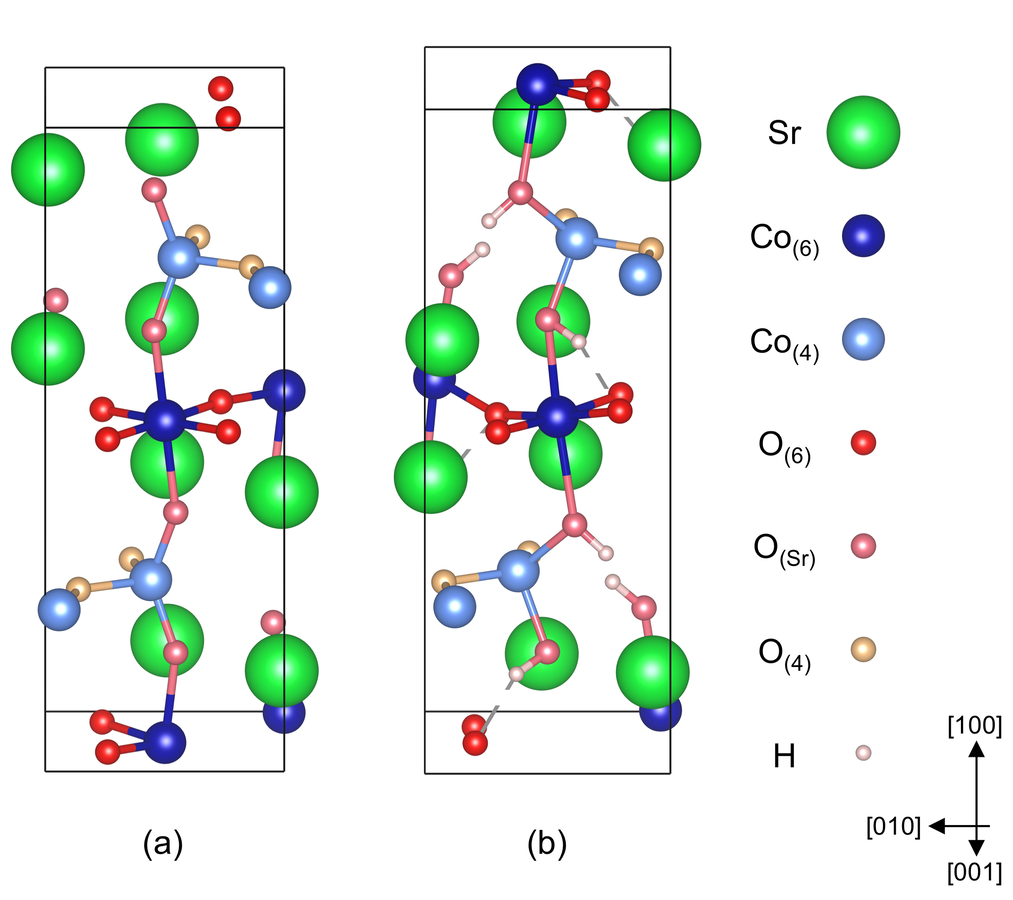}
	\caption{%
		(Color online) 
		illustrations of the (a)~{BM-SCO} and (b)~{H-SCO} phases; 
		the different Co and O sites are here color-coded for illustrative %
		purpose. }
	\label{fig:StructIllust}
\end{figure}

For the further engineering of SCO, especially for fuel cell applications, 
insights in both the actively-researched {BM-SCO}, and the novel and %
	barely-researched {H-SCO} phases, are indispensable. 
However, some fundamental issues remain unresolved.

Firstly, there are uncertainties regarding their ground-state crystalline, %
	electronic, and magnetic structures. 
Though it has been experimentally \cite{MagStructSCO} and numerically %
	\cite{ElectStructAFMSCO,CrysMagStruct} established that {BM-SCO} has a %
	{G-type} antiferromagnetic (AFM) and semiconducting ground state, 
with a crystallographic space-group symmetry variously described as 
\ImaTwo{} \cite{CrysMagStruct,SCOInterstitialH} or \spacegrp{Imma} %
	\cite{CrysMagCharacterSCO},
it must be noted that a definite value of the bandgap has not been provided in %
	literature, 
with discrepancies among the calculated and measured values (Table~%
	\ref{table:LiteratureGapValues}). 
This problem is likely to be aggravated by the experimental conditions, 
the electronic environment (such as the ionic-liquid gating employed in %
	[\onlinecite{TriStateXform}]), 
the presence of defects, secondary phases, and different local configurations, 
as well as the variance in the actual optical transitions measured.
\begin{table}
	\caption{Comparison of some attested bandgap values in {BM-SCO}. }
	\label{table:LiteratureGapValues}
	\begin{ruledtabular}
	\begin{tabular}{ccccc}
		Reference & Study type & Bandgap (eV) & Notes \\ 
		\hline
		\textcite{ElectStructAFMSCO} & 
			Numerical & 0.69\footnotemark[1] & ${U-J=4}$~eV\footnotemark[2] \\
		\textcite{CrysMagStruct} & 
			Numerical & 0.63\footnotemark[1] & \footnotemark[2] \\
		\textcite{TriStateXform} & 
			Numerical & 0.25\footnotemark[1] & ${U-J=4}$~eV\footnotemark[2] \\
		\textcite{SCOInterstitialH} & 
			Numerical & 1.4 & ${U-J=6.5}$~eV\footnotemark[2] \\
		\textcite{ReversalStructSCO} & 
			Experimental & 0.35 & \\
		\textcite{SCOPhotocatalyst} & 
			Experimental & 1.10 & \\
		\textcite{TriStateXform} & 
			Experimental & 2.12 &
	\end{tabular}
	\end{ruledtabular}
	\footnotetext[1]{Estimated from the densities of states (DOSes). }
	\footnotetext[2]{%
		Generalized-gradient approximation (GGA) with on-site Coulomb %
		interaction {(+$U$)}. }
\end{table}

Even less well understood is the {H-SCO} phase. 
The initial study \cite{TriStateXform} indicates a semiconductor with a wider %
	bandgap relative to {BM-SCO}, 
exhibiting weak ferromagnetism (FM);
however, the presented numerical evidence was insufficient towards confirming %
	the experimental observations. 
Moreover, the measured FM is counterintuitive according to the superexchange %
	picture, 
which for an overwhelming majority of magnetic oxide semiconductors predicts %
	intrinsically AFM couplings; 
the possibility exists that the experimentally-observed FM was %
	carrier-mediated and could be attributed to dopants and defects, 
as is usual for FM in semiconductors \cite{FMSemicondOxide,DefectFMDMO}. 
Indeed, a comparison between the two magnetic couplings would have been %
	instructive towards resolving this difficulty, 
but was wanting. 
Also, properly characterizing the ground-state configuration is a prerequisite %
	for further investigations into diffusion mechanisms in {H-SCO}, 
which are invaluable towards potential device applications. 

Another issue is the robustness of the OVCs. 
In semiconductors, vacancies are often disordered, 
and have positive formation energies \cite{CZTSDefect,ZnOODefect,ZnOVacFormE}. 
However, the oxygen-deficient {BM-SCO} phase is variously attested as stable %
	or metastable \cite{ReversalStructSCO,PhaseTransSCO,SCOEarliest} with %
	respect to oxidation to the fully-oxygenated and perovskite-structured SCO %
	({P-SCO}, formula {SrCoO\textsubscript{3}}). 
Furthermore, the OVCs in {BM-SCO} are arranged in a parallel manner consistent %
	with the aforementioned symmetries, 
which define its highly directional \cite{OrientOVC} ionic transport %
	\cite{OxygenDiffusionSCO} and catalytic %
	\cite{SCOPhotocatalyst,ReversibleRedoxSCO} properties. 
The importance of such robust OVCs towards applications is self-obvious, 
but the physical origin of this stability is yet unclear. 

In view of the promising prospect of {BM-SCO} (and by extension, {H-SCO}), 
and the aforementioned issues surrounding the phases, 
we set out to revisit the basic problem of their stabilities. 
\emph{Qualitatively}, the electron-counting model (ECM) %
	\cite{ElectronCounting}, 
widely used in the investigations of surface reconstructions and in stability %
	analyses of semiconducting surface states, 
may offer insights towards our problem. 
Such models and insights can then be \emph{quantitatively} verified by %
	density-functional theoretical (DFT) calculations, 
which allow for the determination of the ground state of the SCO phases, 
and subsequently their electronic structures. 


\section{Electron-counting models}
\label{sect:ECM}

To the best of our knowledge, 
the application of the ECM formalism to the analysis of defects or defect %
	complexes, 
which the OVCs are, 
is previously unheard of. 
However, since the OVCs of {BM-SCO} (and {H-SCO}) can be considered as %
	specially-aligned inner surfaces, 
they should also in principle satisfy ECM. 
Now, the key to ECM is the fulfilling of the octet rule around atoms by %
	coordination, 
where a fractional number of electrons in each of such assumed ``coordinate %
	bonds'' 
(hereafter called ``bonds'', but not in the strict chemical sense) 
is permitted; 
previously, similar treatment has been proposed in the ECM investigation of %
	{TiO\textsubscript{2}} \cite{ModElectronCounting}. 

We proceed to show that {BM-SCO} [Fig.~{\ref{fig:StructIllust}(a)}] satisfies %
	ECM, 
with the following simplifying assumptions. 
Firstly, we assume that both types of Co centers, 
irrespective of their local electronic environments, 
have six surrounding ``lobes'' of charge distributions. 
This assumption suggests that the four-coordinated or ``tetrahedral'' Co %
	center (\Co{4}) can be considered as a distorted ``octahedral'' (i.e. %
	six-coordinated) center (\Co{6}), 
with two dangling bonds in lieu of the missing O atoms. 
We also note the average valence of Co to be three, 
justified by balancing the formal oxidation states of the elements per %
	chemical formula of {BM-SCO}; 
therefore, a Co atom contributes ${3/6}$ electron to each {Co--O} bond. 

Since ECM requires dangling bonds on the cationic centers to be empty %
	\cite{ElectronCounting}, 
for each \Co{4} center, a surplus of one electron is incurred. 
As for each O atom, invariably bonded to two Co centers regardless of its site, 
it gains in total one electron from the two bonds beside its six valence %
	electrons -- 
still one electron short of an octet. 

With the above discussion, 
and noting how \Co{6} and \Co{4} centers are equal in concentration, 
we examine the electronic surplus per formula of {BM-SCO} (\SCO{}) and show %
	electronic balance to indeed be achieved, 
hence stabilizing the OVCs: 
\begin{equation}
	\label{eq:BMSCOECM}
	\underbrace{1\times\left(+2\right)}_{\text{Sr}}
	+\underbrace{0.5\times\left(0\right)}_{\Co{6}}
	+\underbrace{0.5\times\left(+1\right)}_{\Co{4}}
	+\underbrace{2.5\times\left(-1\right)}_{\text{O}}
	=0.
\end{equation}

As for {H-SCO} [Fig.~{\ref{fig:StructIllust}(b)}], 
we can consider two possibilities in assigning a charge state to hydrogen (H), 
which is known to assume different charge states even in the same material, 
such as Si \cite{HydrogenSi}; 
the determination of which is often a point of contention. 
The first possibility is that H is considered nil-valent, 
equivalent to the weak-{O--H} bonds limit. 
This makes the phase very much like {BM-SCO}, 
only with H incorporated via physisorption. 
The charge transfers and valence changes, if any, 
would be small in this case, 
in essence preserving the {BM-SCO} ECM above. 

The other possibility is that H is taken to be monovalent, 
with strong {O--H} bonds; 
this essentially results in a charge transfer to the \Ox{Sr} atoms, to which H %
	atoms are bonded. 
To simplify matters, 
we just take H as an electron donor \cite{SCOInterstitialH}, like Sr. 
Assume for a moment a Co valence of $v$. 
Each {Co--O} bond thus consists of ${v/6}$ electron from Co and 
	${\left(8-v\right)/6}$ electron from O, 
amounting to an excess of ${v/3}$ electron per \Co{4}. 
Each O has an electron deficiency of ${8-6-2\cdot v/6}={\left(6-v\right)/3}$; 
thus to strike electronic balance, this equality needs to be satisfied: 
\begin{align}
	\label{eq:HSCOECM}
	\underbrace{1\times\left(+1\right)}_{\text{H}}
	&+\underbrace{1\times\left(+2\right)}_{\text{Sr}}
	+\underbrace{0.5\times\left(0\right)}_{\text{\Co{6}}} \nonumber\\
	&+\underbrace{0.5\times\left(\frac{v}{3}\right)}_{\text{\Co{4}}}
	+\underbrace{2.5\times\left(\frac{v-6}{3}\right)}_{\text{O}}
	=0.
\end{align}
Hence we have ${v=2}$, 
meaning that the charge transfer induces a valence change in Co from {(III)} %
	to {(II)}, 
consistent with previous proposals \cite{TriStateXform,SIOHydrogenation}; 
which of the pictures is more descriptive of {H-SCO} remains to be verified by %
	calculations. 


\section{Computational details}
\label{sect:ComputDetails}


\subsection{General methodology}
\label{sect:GeneralMthd}

The Vienna \emph{Ab-initio} Simulation Package (VASP) \cite{vasp1,vasp2} was %
	used to perform all calculations. 
Spin-polarized calculations were done using the Perdew--Burke--Ernzerhof %
	formulation of the generalized-gradient approximation (PBE-GGA) %
	\cite{PBE-GGA,*PBE-GGAErrata}, 
with pseudopotentials (PPs) generated by the projector augmented-wave (PAW) %
	\cite{PAW,PAW-PP} method. 
A {G-type} spin texture was assumed for AFM calculations. 
Plane-wave bases were truncated to a {400-eV} energy cutoff, 
and gamma-centered {k-point} meshes were used to sample the reciprocal space. 
All lattice-vector, positional, and electronic degrees of freedom were relaxed, 
with forces on individual atoms converged to under 0.02~{eV/\Ang{}}. 
For atomic charge [resp. magnetic moment (MM)] analyses, 
the integrated site- and orbital-projected charge (resp. spin) density around %
	each atom was found by the PAW-based quick projection scheme as %
	implemented in VASP. 

\begin{figure}
	\includegraphics[width=3.375in]{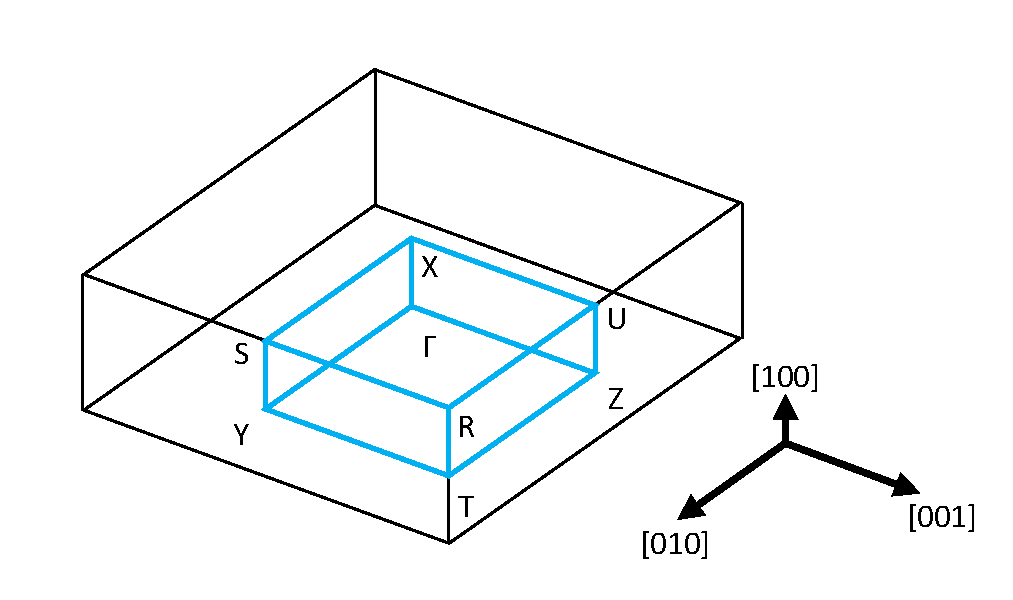}
	\caption{%
		Schematics of the first BZ (in thin lines) and the IBZ (in thicker %
			lines) of the {BM-SCO} and {H-SCO} phases, 
		with the high-symmetry points labeled. }
	\label{fig:BZ}
\end{figure}

Simulation cells were constructed based on a fully-relaxed {36-atom} {BM-SCO} %
	primitive cell, 
similar to the cell of \ImaTwo{} symmetry described by {Mu\~noz} \emph{et al.} %
	\cite{CrysMagStruct} 
(It should be noted that the reported {G-type} AFM necessitated the doubling %
	of the primitive cell.) 
Single-H adsorption was studied using a {144-atom} $1\times2\times2$ %
	{BM-SCO} host supercell. 
The {k-point} meshes used for primitive-cell and supercell calculations were %
	$2\times5\times5$ and $2\times3\times3$ respectively; 
after full relaxations, band structures were calculated by sampling {k-points} %
	along the twelve edges of the irreducible Brillouin zone (IBZ) (Fig.~%
	\ref{fig:BZ}) 
	of the orthogonal primitive cell. 
The {k-point} mesh size, and the plane-wave energy cutoff, 
were convergence-tested to ensure the physicality of the results.

The effects of correlation between {Co-\orb{d}{}} electrons were included by %
	considering the on-site Coulomb interaction ($U$), 
utilizing the simplified, rotationally-invariant Hubbard model %
	\cite{HubbardModel} by Dudarev \emph{et al.} \cite{LDAUTYPE2}, 
where the Coulomb $U$ and the exchange $J$ parameters enter the Hamiltonian %
	via only their difference ${U-J}$, 
which we simplified to $U$ by choosing ${J=0}$. 
After extensive tests (see Section~\ref{sect:UFuncTest}), 
we settled on choosing ${U=4}$~eV, 
in keeping with previous practice in {BM-SCO} calculations %
	\cite{TriStateXform,CrysMagStruct}. 


\subsection{\checkboldmath{U} and functional testing}
\label{sect:UFuncTest}

Seeing how {H-SCO} and {BM-SCO} should be \emph{a priori} treated as %
	distinct materials, 
and established parameters which work for {BM-SCO} may be suboptimal for %
	{H-SCO}, 
we conducted tests on the structural, electronic, and magnetic properties of %
	{H-SCO}, 
with respect to the choice of exchange--correlation functionals and $U$ %
	parameters. 
Based on the configuration reported in [\onlinecite{TriStateXform}], 
we constructed a symmetrized {H-SCO} unit cell %
	[Fig.~{\ref{fig:StructIllust}(b)}]; 
which we then relaxed using the local-density-approximation (LDA) and PBE-GGA %
	functionals, 
in conjunction with their corresponding PP sets, 
with the $U$ parameter set between 0 and 5~eV. 

\begin{figure}
	\includegraphics[width=3.375in]{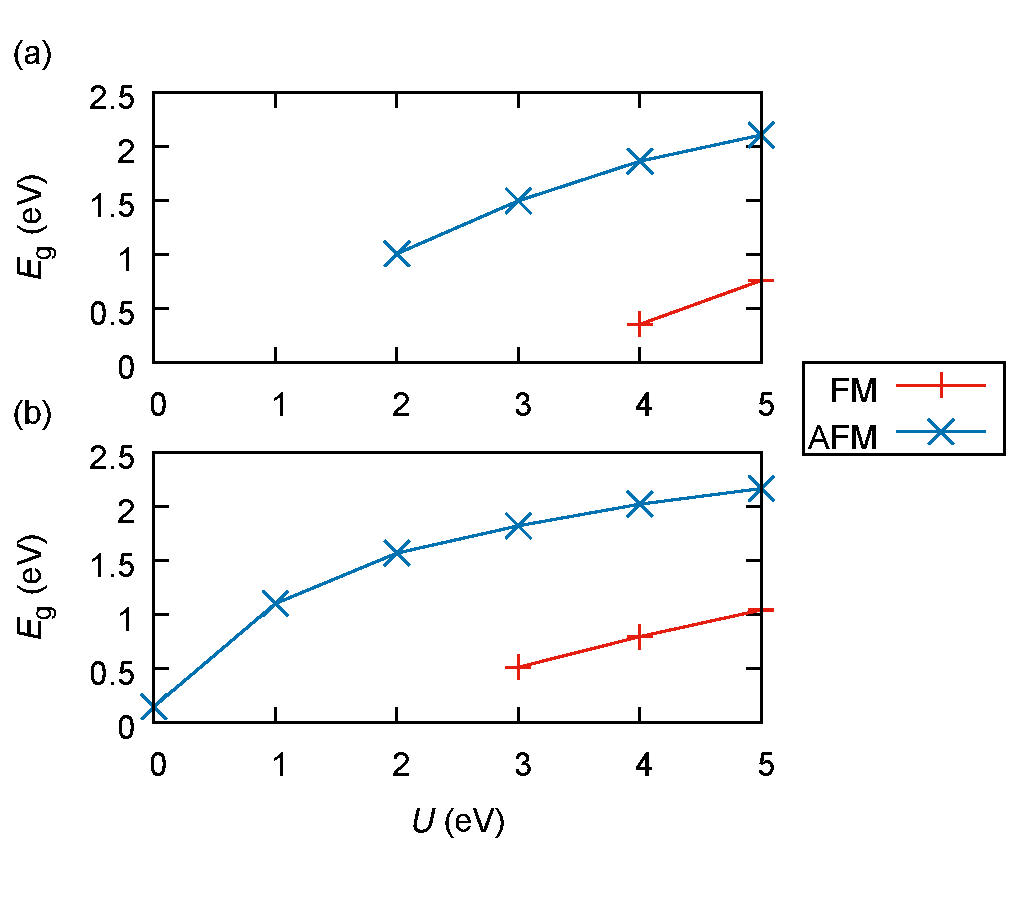}
	\caption{%
		Plots of the bandgap energy $E_{\text{g}}$ of {H-SCO}, if any, 
		against the $U$ parameter, 
		for the (a)~LDA and (b)~PBE-GGA functionals; 
		bandgaps estimated from the eigenvalues of the highest-occupied and %
			lowest-unoccupied bands on the $2\times5\times5$ {k-point} mesh. }
	\label{fig:UGap}
\end{figure}

The trend for the bandgap energy is consistent across both functionals %
	(Fig.~\ref{fig:UGap}) that it increases with increasing $U$; 
and for LDA, the gap closes when $U$ is small. 
In view of the prominent {Co-\orb{d}{}} character of the valence and %
	conduction bands of {BM-SCO} \cite{CrysMagStruct}, 
we would expect the situation to be somewhat similar in {H-SCO}. 
Since the Hubbard $U$ term adjusts the strength of the on-site Coulomb %
	interaction between {Co-\orb{d}{}} electrons, 
the band edges (and thus the bandgap energy) are naturally highly dependent on %
	it. 
Due to the loss of the bandgap with LDA at small $U$, 
we deemed such settings to be unsuitable for the description of {H-SCO}, 
although they gave good agreement to the crystalline structure (Supp. Fig.~2) %
	\cite{Note1}.

\begin{figure}
	\includegraphics[width=3.375in]{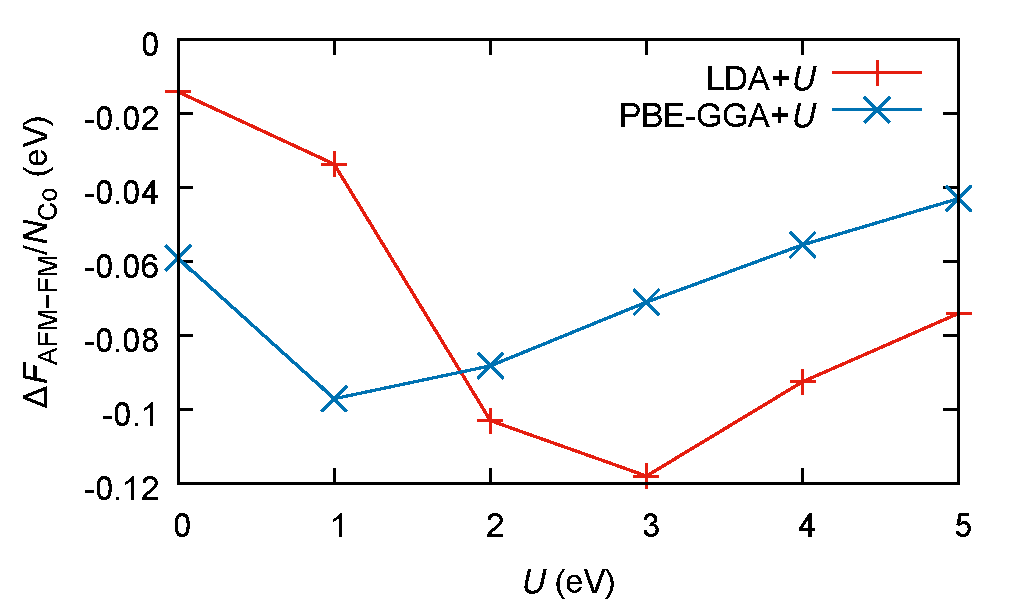}
	\caption{%
		Plot of the magnetic coupling energy \MagE{} of {H-SCO} against the %
			$U$ parameter, 
		for both the LDA and PBE-GGA functionals. } 
	\label{fig:UMag}
\end{figure}

We also tested for the magnetic coupling energy \MagE{}, 
i.e. the free energy difference per Co atom between the AFM and FM phases, 
to check the energetic favorability of the two couplings. 
To our surprise, AFM was found to be energetically favorable relative to FM, 
consistently for both functionals (Fig.~\ref{fig:UMag}) at all tested values %
	of $U$, 
in apparent contradiction to the experimental observation in %
	[\onlinecite{TriStateXform}]. 


\section{Results and discussions}
\label{sect:Results}

[Rf. SM \cite{Note1} for band-component analyses (Section~IV, SM) %
	and \texttt{.cif} files (Section~V, SM) of the three discussed {BM-SCO} %
	structures, 
namely, \emph{phases 1--3}, and that of the lowest-energy {H-SCO} structure.]


\subsection{The brownmillerite phase}
\label{sect:ResultsBMSCO}

\begin{figure}
	\includegraphics[width=3.375in]{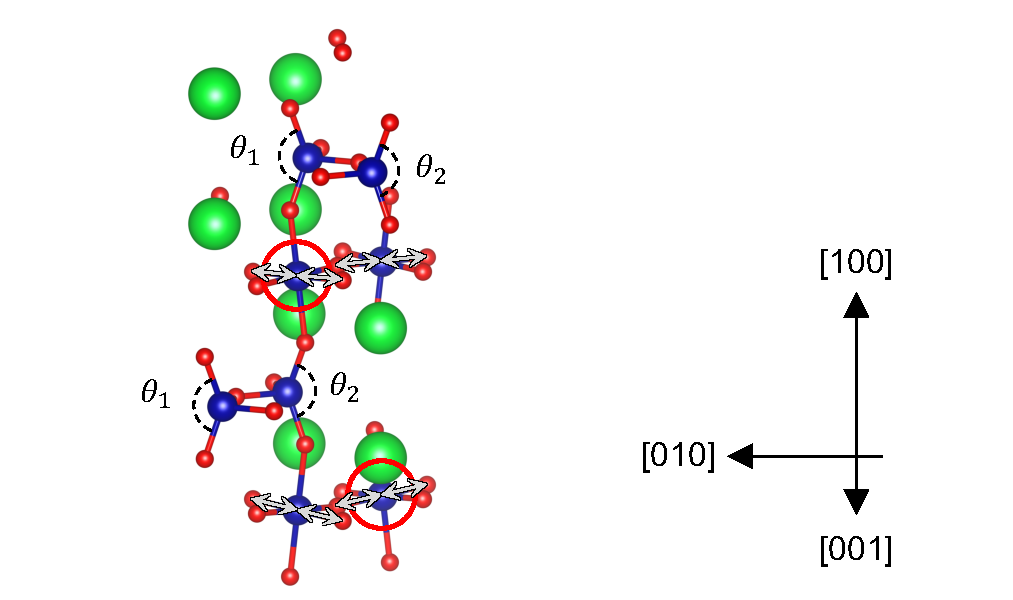}
	\caption{%
		Illustration of the distribution of the longer {\Co{6}--\Ox{6}} bonds 
		(overlaid with gray arrows) 
		in \change{\emph{phase-1} {BM-SCO}; 
		note how the bond arrangements about the two circled \Co{6} centers, 
		$\left(\frac{1}{2},\frac{1}{2},\frac{1}{2}\right)$ apart in direct %
			coordinates, 
		are different. }
		Also shown: 
		definitions of the {\Ox{Sr}--\Co{4}--\Ox{Sr}} bond angles $\theta_1$ %
			and $\theta_2$. }
	\label{fig:BMSCOAngles}
\end{figure}

\begin{figure*}
	\includegraphics[width=6.75in]{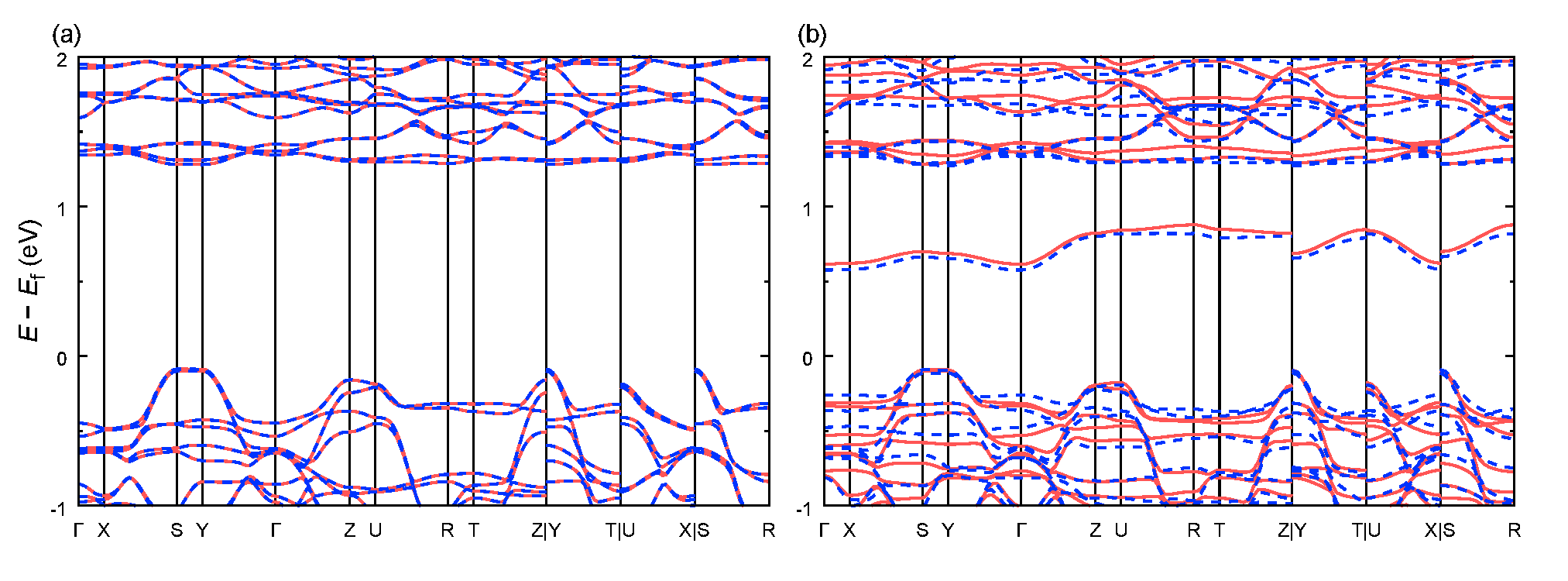}
	\caption{%
		(Color online) 
		plots of the band structures of (a) \emph{phase-1} and (b) %
			\emph{phase-2} {BM-SCO} near the band edges, 
		with spin-up (resp. -down) bands in red (color)/light gray %
			(grayscale) solid lines [resp. blue (color)/dark gray (grayscale) %
			dotted lines]. }
	\label{fig:BMSCOBands}
\end{figure*}

A {BM-SCO} cell was first obtained after detailed structure optimization. 
The most stable structure we obtained, hereafter referred to as \emph{phase~1}, 
retains from the \ImaTwo{} space group the reflectional symmetry about the %
	OVCs, 
and the \spacegrp{2\textsubscript{1}} screw axes along them. 
However, a split among the {\Co{6}--\Ox{6}} bond lengths to %
	${2.052\!\left(4\right)}$~\Ang{} and ${1.84660\!\left(8\right)}$~\Ang{} %
	\change{eliminates the translational symmetry between the two planes of %
		\Co{6} (Fig.~\ref{fig:BMSCOAngles}), 
	thus lowering} the symmetry to a shifted \PmcTwoSubOne{}. 
It has large direct bandgaps of 1.37~eV (Table~\ref{table:BMSCOPhases}) at the %
	Y and S points [Fig.~{\ref{fig:BMSCOBands}(a)}], 
with the valence and conduction bands (VB and CB, resp.) governed by %
	the coupling between {Co-3\orb{d}{}} and {O-2\orb{p}{}} orbitals (Supp. %
	Fig. 8)\cite{Note1}. 
\begin{table}
	\caption{Comparison of the three obtained {BM-SCO} phases, 
		with \emph{phase 1}, \emph{2}, and \emph{3} being %
		\PmcTwoSubOne{}-orthorhombic, slightly monoclinic, and %
		\ImaTwo{}-orthorhombic, respectively. }
	\label{table:BMSCOPhases}
	\newlength\myw                          
	\setlength{\myw}{.22\columnwidth}       
	\begin{ruledtabular}
	\begin{tabular}{p{.3\columnwidth}ccc}
		\multicolumn{1}{c}{Property} & \emph{Phase~1} & \emph{Phase~2} & 
			\emph{Phase~3} \\ 
		\hline
		Lattice constants $a$, $b$, $c$ (\Ang{}) & 
			\parbox[t]{\myw}{\centering 15.589, 5.570, 5.438} & 
			\parbox[t]{\myw}{\centering 15.636, 5.334, 5.440\footnotemark[1]} &
			\parbox[t]{\myw}{\centering 15.529, 5.601, 5.401} \\
		{\Ox{Sr}--\Co{4}--\Ox{Sr}} angle $\theta_1$ & 
			139\dgsign{} & 138\dgsign{} & 126\dgsign{} \\
		{\Ox{Sr}--\Co{4}--\Ox{Sr}} angle $\theta_2$ & 
			139\dgsign{} & 141\dgsign{} & 126\dgsign{} \\
		Relative free energy ${\Delta F}$ per formula (eV) & 
			0 & +0.123 & +0.0368 \\
		\Co{6} MM (\BohrM{}) & 
			2.959 & ${2.971\!\left(2\right)}$ & 2.958 \\
		\Co{4} MM (\BohrM{}) & 
			2.915 & ${2.4\!\left(5\right)}$ & 
			$2.907\!\left(1\right)$ \\
		Bandgap energy (eV), nature & 
			\parbox[t]{\myw}{\centering 1.37, direct} & 
			\parbox[t]{\myw}{\centering 0.67, indirect} & 
			\parbox[t]{\myw}{\centering 0.93, (estimate)\footnotemark[2]}
	\end{tabular}
	\end{ruledtabular}
	\footnotetext[1]{Indexed with reference to the \emph{phase-1} cell. }
	\footnotetext[2]{Estimated from the eigenvalues sampled on the {k-point} %
		mesh. }
\end{table}

In addition to this most-stable \emph{phase~1}, 
we also obtained an almost-orthorhombic ($90\dgsign{} < \alpha < %
	90.1\dgsign{}$) phase with a relatively small gap, 
which we named \emph{phase~2}; 
this phase is 123~meV higher in energy compared to \emph{phase~1} for each %
	formula of \SCO{}. 
The bandgap energy also decreases to 0.67~eV, 
with the CB minimum (CBM) shifting to the $\Gamma$~point [Fig.~%
	{\ref{fig:BMSCOBands}(b)}]. 
[Note that as an approximation, the reciprocal space of \emph{phase-2} %
	{BM-SCO} is traversed (Fig.~\ref{fig:BZ}) as if the cell is orthorhombic.]
Beside the unit-cell distortion, some other geometric differences between %
	these two structures are noted, 
mainly in the {\Ox{Sr}--\Co{4}--\Ox{Sr}} bond angles (Fig.~%
	\ref{fig:BMSCOAngles}): 
in \emph{phase~1}, the angles ${\theta_1\approx\theta_2\approx139\dgsign{}}$; 
while for \emph{phase~2}, ${\theta_1\approx141\dgsign{}}$ and %
	${\theta_2\approx138\dgsign{}}$, 
breaking the \spacegrp{2\textsubscript{1}}-screw axis symmetry along the OVCs. 
Such local structural distortions induce the decrease of the \Co{4} MMs %
	(Table~\ref{table:BMSCOPhases}), 
creating also a predominantly {\Co{4}--\Ox{Sr}} gap state [Supp. %
	Fig.~{9(b),~(e)}] \cite{Note1}, 
causing a smaller overall bandgap; 
intuitively, the {\orb{p}{}--\orb{d}{}} couplings are changed upon the %
	symmetry-breaking. 

As a control, another orthorhombic phase, \emph{phase 3}, 
was obtained by relaxation under a forced \ImaTwo{} \cite{CrysMagStruct} %
	symmetry. 
Despite the minute differences in the lattice parameters (Table~%
	\ref{table:BMSCOPhases}), 
the bandgap width decreases to 0.93~eV; 
the structure is also 36.8~meV per formula of BM-SCO higher in energy than %
	\emph{phase 1}. 
\change{Moreover, our phonon calculations\cite{Note1} also indicate the %
	dynamical instability of this \ImaTwo{} phase with respect to the %
	\PmcTwoSubOne{} \emph{phase 1}.}
As the difference in space groups arises from the differing Co--O bond lengths %
	on the octahedral layer (Fig.~\ref{fig:BMSCOAngles}), 
the spontaneous and energetically-favorable symmetry-breaking in going from %
	\emph{phase 3} to \emph{phase 1} can be seen as a manifestation of the %
	Jahn--Teller effect \cite{JahnTeller1,JahnTeller2}. 

The above results also suggest how sensitive the electronic and magnetic %
	properties of {BM-SCO} are to the local structures, 
especially near the OVC; 
the phases, which can all be considered perturbations to a predominantly %
	\ImaTwo{} symmetry, 
may easily interconvert due to thermal fluctuations at typical working %
	temperatures of solid oxide fuel cells (SOFCs), 
which are in the order of $10^2$ to $10^3$~K \cite{LowTempFuelCell}. 
This may also explain why previous theoretical studies %
	predicted small bandgaps, 
inconsistent with recent experiments (Table~\ref{table:LiteratureGapValues}). 


\subsection{The hydrogenated phase}
\label{sect:ResultsHSCO}


\subsubsection{Lowest-energy configuration}
\label{sect:ResultsHConfigs}

\begin{figure}
	\includegraphics[width=3.375in]{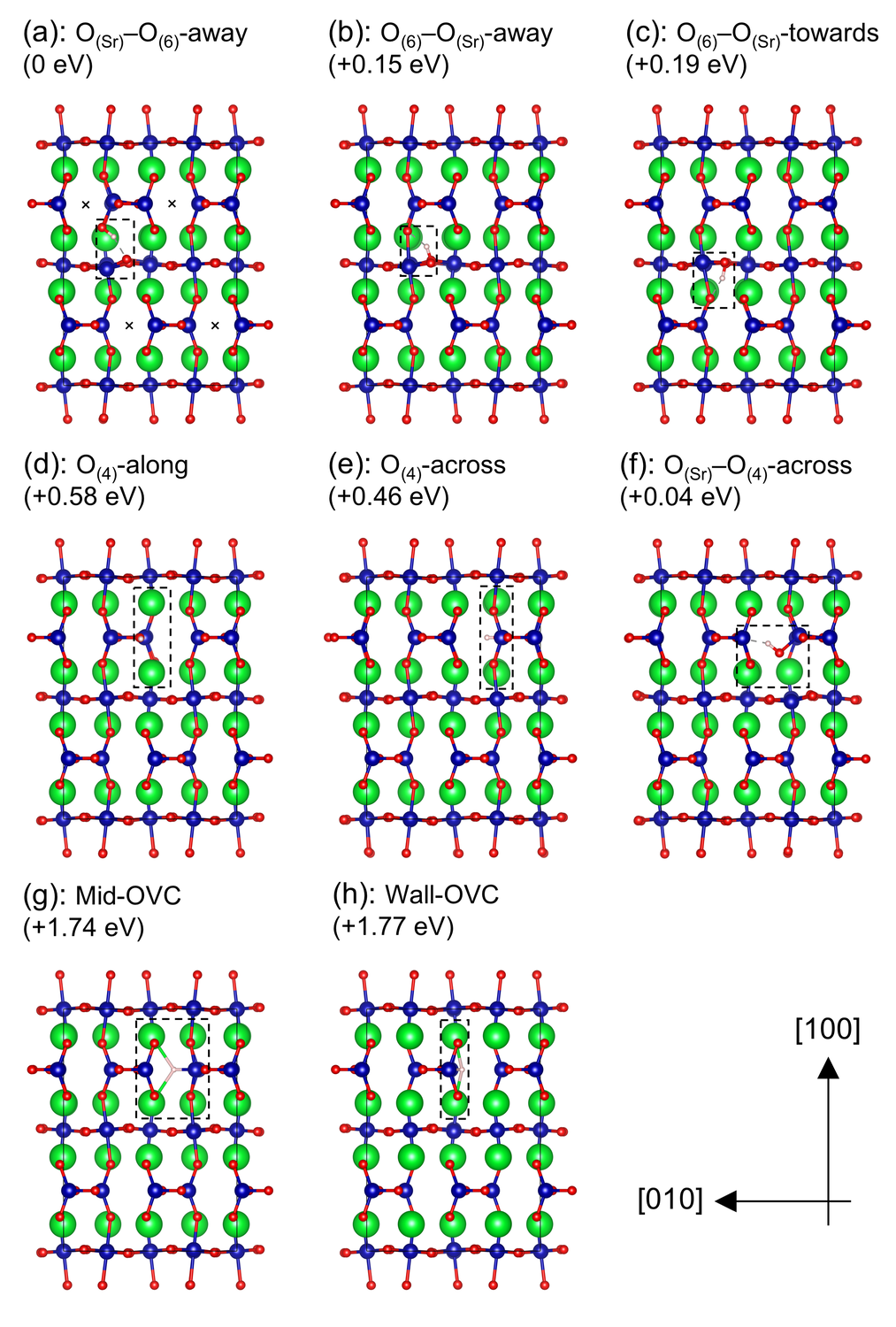}
	\caption{%
		Single-H adsorption sites, 
		viewed along the ${\left[00\bar{1}\right]}$ direction; 
		adsorbed H and surrounding atoms boxed as visual aid. 
		Relative free energies ${\Delta F}$ of the $1\times2\times2$ %
			supercells are also listed above each figure. 
		OVCs are marked with crosses in (a). }
	\label{fig:1HConfigs}
\end{figure}

To gain insights towards the construction of a low-energy {H-SCO} phase, 
we first examined the preferred adsorption sites at the dilute limit. 
A single H atom was introduced to the large ($1\times2\times2$) {BM-SCO} %
	supercell at different possible adsorption sites, 
which are listed in Fig.~\ref{fig:1HConfigs} along with their relative free %
	energies. 
The most energetically-favorable configuration, 
``\Ox{Sr}--\Ox{6}-away'' [Fig.~{\ref{fig:1HConfigs}(a)}], 
has the H atom bonded to an \emph{\Ox{Sr}}, 
pointing \emph{away} from the OVC towards a nearby \emph{\Ox{6}}. 
This differs from earlier studies \cite{TriStateXform,SCOInterstitialH} %
	suggesting the relative stability of the ``\Ox{Sr}--\Ox{4}-across'' %
	configuration [Fig.~{\ref{fig:1HConfigs}(f)}] at the dilute limit, 
which we found to be marginally less favorable energetically by 40~meV; 
relative to the ``adsorption--coordination site'' 
(i.e. \Ox{6} [Fig.~{\ref{fig:1HConfigs}(b),~(c)}], \Ox{Sr} [Fig.~%
	{\ref{fig:1HConfigs}(a),~(f)}], and between metal ions [Fig.~%
	{\ref{fig:1HConfigs}(d),~(e)}]) 
on which the H atom settle, 
its orientation would seem to exert a much smaller effect on the energy. 

Other configurations, like having the H atom directly adjacent to a Co site, 
	were also sampled, 
but they were unfeasible and eventually converged to the listed ones. 
Still, we do note that it is almost impossible to exhaust all the possible %
	configurations, 
given the multitudes of adsorption sites and H orientations, 
even when we only considered the {single-H} cases; 
analysis of the intermediate-hydrogenation regime would be notedly %
	\cite{SCOInterstitialH} and prohibitively difficult. 

\begin{figure}
	\includegraphics[width=3.375in]{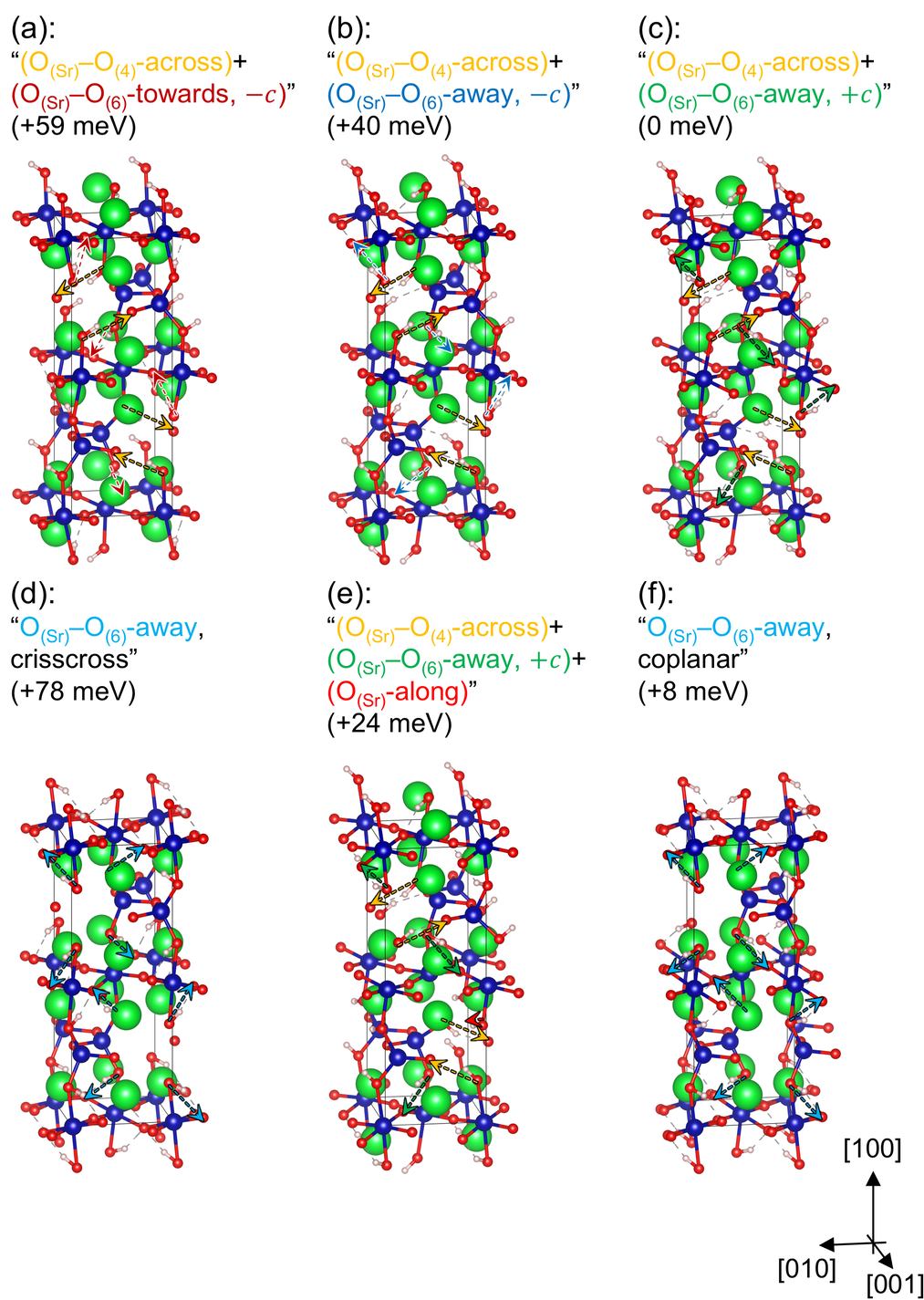}
	\caption{%
		(Color online) 
		fully-hydrogenated {H-SCO} primitive cells, 
		and their relative free energies ${\Delta F}$ per formula of \HSCO{}. 
		The eight arrows in each subfigure indicate the approximate %
			orientations of the {\Ox{Sr}--H} bonds. } 
	\label{fig:FullHConfigs}
\end{figure}

As discussed in Section~III, SM \cite{Note1}, we have shown the {single-H} 
	states to be semiconducting and localized, 
which suggests the generality of the stable local configurations of %
	adsorbed H atoms, 
and their relevance even at elevated H concentrations. 
Hence, based on the {single-H} results, 
we then proceeded to construct fully-hydrogenated {H-SCO} unit cells by %
	putting H atoms onto \Ox{Sr}, 
which is the preferred site at the dilute limit, 
when also ideally preserving the \ImaTwo{}-like symmetry of {BM-SCO}. 
Among the tested configurations (Fig.~\ref{fig:FullHConfigs}), 
``{(\Ox{Sr}--\Ox{4}-across)}+{(\Ox{Sr}--\Ox{6}-away,~${+c}$)}'' [Fig.~%
	{\ref{fig:FullHConfigs}(c)}], 
i.e. that which was constructed for the functional and $U$-parameter tests %
	[Fig.~{\ref{fig:StructIllust}(b)}], 
has the lowest energy. 
This configuration has H orientations arranged highly symmetrically, 
with half the H pointing \emph{across} the OVCs to opposing \emph{\Ox{4}}; 
and half towards \emph{\Ox{6}} pointing \emph{away} from the OVCs, 
in the ${+c}$ direction relative to the adsorbing \Ox{Sr}, 
essentially mixing the lowest-energy {single-H} configurations reported in %
	previous investigations \cite{TriStateXform,SCOInterstitialH} and ours in %
	equal proportions. 
Such is identical to the upper channel in the configuration presented in %
	[\onlinecite{TriStateXform}], 
which we also attempted to replicate; 
however, it settled into the ``{(\Ox{Sr}--\Ox{4}-across)}+%
	{(\Ox{Sr}--\Ox{6}-away,~${+c}$)}+{(\Ox{Sr}-along)}'' configuration [Fig.~%
	{\ref{fig:FullHConfigs}(e)}], 
qualitatively differing from the lowest-energy configuration [Fig.~%
	{\ref{fig:FullHConfigs}(c)}] only by turning one of the four H atoms in %
	the ``{\Ox{Sr}--\Ox{4}-across}'' orientation such that it now points %
	\emph{along} the OVC. 
This break of symmetry is however energetically unfavorable by a moderate %
	margin of 24~meV per formula.

Other configurations were formed by orienting the {\Ox{Sr}--H} bonds towards %
	different \Ox{6} atoms [Fig.~{\ref{fig:FullHConfigs}(a),~(b)}], 
such that the {O--H} bonds point \emph{towards} [Fig.~%
	{\ref{fig:FullHConfigs}(a)}] or \emph{away} from [Fig.~%
	{\ref{fig:FullHConfigs}(b)}] the OVCs, 
in the ${-c}$ direction; 
these are found to incur considerable energy penalties. 
Intriguingly, pointing all {O--H} bonds \emph{away} from the OVCs [Fig.~%
	{\ref{fig:FullHConfigs}(d)}] like the {single-H} favored case [Fig.~%
	{\ref{fig:1HConfigs}(a)}] in a \emph{crisscross} manner turns out to be %
	very energetically unfavorable; 
unless one aligns all {O--H} bonds to be \emph{coplanar} along the (011) plane %
	[Fig.~{\ref{fig:FullHConfigs}(f)}], 
inducing a shrink along said diagonal upon cell relaxation and resulting in a %
	monoclinic cell with $\alpha\approx84\dgsign{}$, 
which is relatively low in energy. 
This is reminiscent of the polymorphism of crystal symmetries found in %
	perovskites \cite{PerovGeom}, 
to which brownmillerites (of which {H-SCO} is a derivative) are closely %
	related. 


\subsubsection{Structural, electronic, and magnetic properties}
\label{sect:ResultsHSCOProps}

The lattice constants $a$, $b$, and $c$ in the obtained {H-SCO} phase increase %
	to 16.064~\Ang{}, 5.727~\Ang{}, and 5.626~\Ang{} respectively, 
indicating lattice expansion relative to {BM-SCO} due to hydrogen %
	incorporation. 
The primitive cell exhibits a shifted \PnaTwoSubOne{} symmetry, 
preserving the \spacegrp{2\textsubscript{1}} screw axes of the brownmillerite %
	structure along the OVCs and the glide planes (100) and (200) of the %
	idealized \ImaTwo{} cell, 
at the cost of the reflection symmetry of the OVCs about the {\Co{4}--\Ox{4}} %
	planes. 
This is in contrast with the \PmcTwoSubOne{} {BM-SCO} cells, 
where the glide planes are lost instead. 
Indeed, the change of space groups upon hydrogenation is not unheard %
	of; 
as a more drastic example, hydrogen-doping proves effective in stabilizing the %
	zincblende phase against the wurtzite phase in, 
for example, zinc oxide and gallium nitride \cite{ImpurPhaseStablSemicond}. 

\begin{figure}
	\includegraphics[width=3.375in]{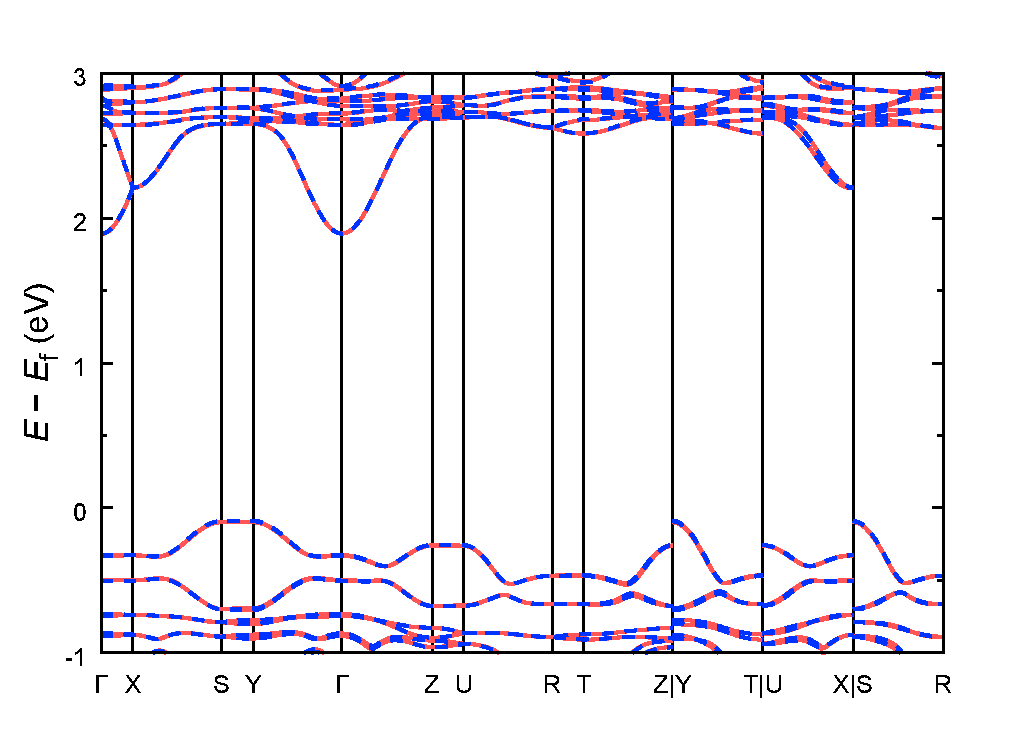}
	\caption{%
		(Color online) 
		plot of the band structure of {H-SCO} near the band edges, 
		with spin-up (resp. -down) bands in red (color)/light gray %
			(grayscale) solid lines [resp. blue (color)/dark gray (grayscale) %
		dotted lines]. }
	\label{fig:HSCOBands}
\end{figure}

The change in the electronic structure (Fig.~\ref{fig:HSCOBands}) relative to %
	the {BM-SCO} phase is drastic. 
{H-SCO} has a bandgap {1.99-eV} wide, 
620~meV wider than that of {BM-SCO}, 
in qualitative agreement with previous results \cite{TriStateXform} that the %
	bandgap widened by 720~meV upon hydrogenation. 
Like in {BM-SCO}, the valence and conduction bands are still heavy in %
	{Co-3\orb{d}{}} and {O-2\orb{p}{}} characters, 
yet complications arise: 
the CBM is now shifted to the $\Gamma$~point, 
acquiring a somewhat {\orb{s}{}--\orb{d}{}} character (Supp. Fig.~10) %
	\cite{Note1}.  
Compared to the {BM-SCO} CBM, which is of {\orb{p}{}--\orb{d}{}} character, 
the band curvature is much increased, 
suggesting enhanced carrier mobility when the system is lightly %
	electron-doped. 

As mentioned in Section~\ref{sect:UFuncTest}, 
in seeming contradiction with the experimentally-reported FM %
	\cite{TriStateXform}, 
the {G-type} AFM in {BM-SCO} is preserved, 
and is stabilized against FM by 56~meV per Co atom. 
The finding is also qualitatively supported by hybrid and meta-GGA DFT %
	results ({Section~I~C}, SM) \cite{Note1}. 
This apparent inconsistency, as we will later address in Section~%
	\ref{sect:HoleFM}, 
may possibly be attributed to hole-mediated FM in the experimental sample. 
The MMs on \Co{6} and \Co{4} become 2.627~\BohrM{} and 2.647~\BohrM{} %
	respectively, 
about 0.3~\BohrM{} lower compared to those in {BM-SCO}.

\begin{figure}
	\includegraphics[width=3.375in]{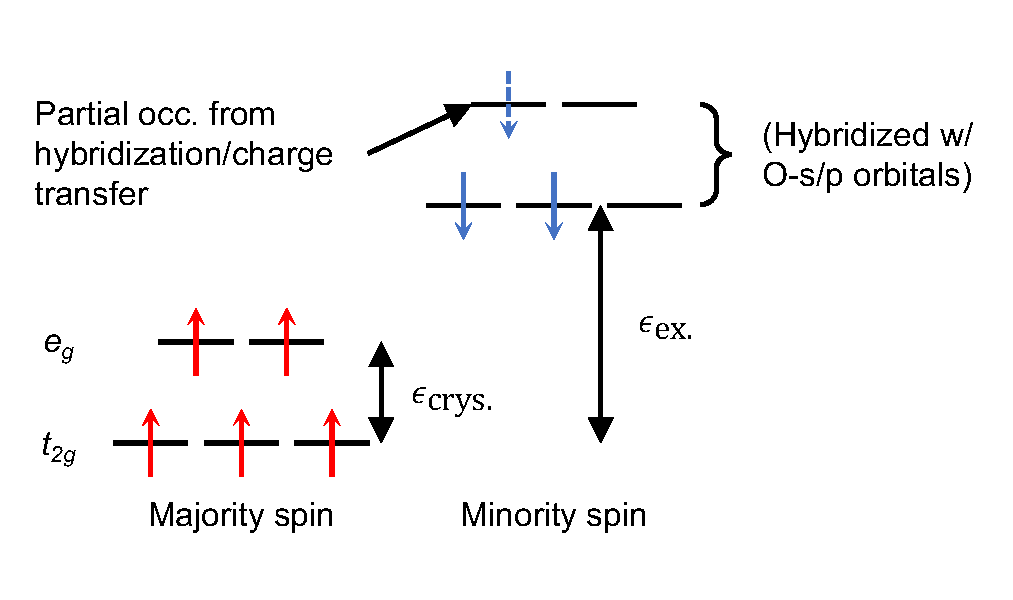}
	\caption{%
		Schematics of the energy splitting of the Co-\orb{d}{} orbitals. }
	\label{fig:dSplitting}
\end{figure}

From Table~\ref{table:ProjDensities}, 
it is seen that the projected charge and MMs of the {Co-\orb{d}{}} orbitals %
	are qualitatively similar, 
consistent with our ECM where \Co{6} and \Co{4} centers are treated as having %
	the same valence, 
and identical up to the number of O ligands. 
We hereby give a simplistic argument towards the Co MM, 
in terms of the atomic orbital couplings; 
we begin by considering {BM-SCO}. 
Assume for Co a \orb{d}{}\textsuperscript{7} configuration (Fig.~%
	\ref{fig:dSplitting}). 
Considering the assumed \spacegrp{O\textsubscript{H}} symmetry of a Co center, 
the \spacegrp{e\textsubscript{g}} (\orb{d}{z^2}, \orb{d}{x^2\!-\!y^2}) %
	orbitals are split upwards relative to the \spacegrp{t\textsubscript{2g}} %
	(\orb{d}{xy}, \orb{d}{yz}, \orb{d}{zx}) orbitals by the crystal field. 
With sufficiently large exchange splitting $\epsilon_{\text{ex.}}$ between the %
	spins, 
relative to the crystal-field splitting $\epsilon_{\text{crys.}}$, 
the five majority-spin states would be fully occupied, 
and the remaining electrons occupy two of the minority-spin %
	\spacegrp{t\textsubscript{2g}} states.
\begin{table}
	\caption{Select orbital projections of the charges and magnetic moments %
		around various species in {BM-SCO} and {H-SCO}. }
	\label{table:ProjDensities}
	\begin{ruledtabular}
	\begin{tabular}{ccccc}
		\multirow{2}{*}{Orbital} & \multicolumn{2}{c}{Charge (elem. charge)} & 
			\multicolumn{2}{c}{MM (\BohrM{})} \\ \cline{2-5}
		 & {BM-SCO} & {H-SCO} & {BM-SCO} & {H-SCO} \\ 
		\hline
		{\Co{6}-\orb{d}{}} & 
			6.982 & 7.173 & 2.955 & ${2.6055\!\left(5\right)}$ \\
		{\Co{4}-\orb{d}{}} & 
			${7.0865\!\left(5\right)}$ & ${7.1695\!\left(5\right)}$ & 2.863 & 
				2.591 \\
		{\Ox{Sr}-\orb{d}{}} & 
			3.439 & ${3.555\!\left(8\right)}$ & ${0.1595\!\left(5\right)}$ & 
				${0.022\!\left(4\right)}$ \\
		{H-\orb{s}{}} & 
			N/A & ${0.606\!\left(3\right)}$ & N/A & ${0.0005\!\left(5\right)}$
	\end{tabular}
	\end{ruledtabular}
\end{table}

Now, we consider the hopping of electrons between the Co and O orbitals, 
as described in the tight-binding and Hubbard \cite{HubbardModel} models. 
For convenience, we align the {$z$-axis} along the {Co--O} ``bond''. 
By symmetry arguments, 
we see that the {Co-\orb{d}{z^2}} (i.e. \spacegrp{e\textsubscript{g}}) and %
	{O-\orb{p}{z}} (or {O-\orb{s}{}}) orbitals directly overlap, 
allowing the electrons to easily hop between the orbitals. 
(This is, in the hybridization language, {$\sigma$-bonding}.) 
Less direct yet still permitted is hopping (i.e. {$\pi$-bonding}) between the %
	{Co-\orb{d}{xy}} (i.e. \spacegrp{t\textsubscript{2g}}) and {O-\orb{p}{x}} %
	orbitals, 
and the {Co-\orb{d}{yz}} and {O-\orb{p}{y}} orbitals. 
This creates further splitting among the {Co-\orb{d}{}} states, 
partially populating the minority-spin {Co-\orb{d}{}} levels with electrons %
	which hopped from O, 
thus contributing to a \orb{d}{}-orbital MM of slightly less than %
	three.

As for {H-SCO}, it is seen that the {O--H} bond strength is at a delicate %
	balance which gives rise to the interesting and useful properties of %
	{H-SCO} seen experimentally. 
Around H atoms, the charge is qualitatively close to unity ($\approx{0.6}$~%
	elem. charge) (Table~\ref{table:ProjDensities}), 
and the increase of that around \Ox{Sr} is small ($\approx{0.1}$~elem. charge) %
	compared to the \Ox{Sr} charge in {BM-SCO}. 
This indicates a weak charge transfer from H to \Ox{Sr}, 
evident of physisorption of H in a {BM-SCO} scaffold, 
corresponding to the proposed weakly-bonded ECM (rf. Section~\ref{sect:ECM}). 
However, we also note the charge in the {Co-\orb{d}{}} orbitals to increase %
	somewhat ($\approx{0.1}$~elem. charge), 
hinting at {H--\Ox{Sr}--Co} hopping of the partially-transferred charge, 
which is qualitatively consistent with the slight decrease ($\approx{0.3}$~%
	\BohrM{}) in the {Co-\orb{d}{}} MMs. 
By virtue of this small increase in the occupation of the hopping-allowed %
	{Co-\orb{d}{}} orbitals by {H-\orb{s}{}} electrons, 
hydrogenation suppresses the {Co-\orb{d}{}}-to-{O-\orb{p}{}} coupling as %
	previously suggested \cite{TriStateXform}, 
which affects the energy splitting and ordering of the bands, 
resulting in the exposure of the new CBM of increased {\orb{s}{}--\orb{d}{}} %
	character (Supp. Fig.~10) \cite{Note1}. 
Such suppressed {\orb{p}{}--\orb{d}{}} coupling is consistent with the %
	increase in the charge density around Co, 
and decreases in Co valence and magnetization, 
corresponding to the strongly-bonded ECM proposed towards the end of Section~%
	\ref{sect:ECM}. 
As such, {H-SCO} is best considered to be at the middle-ground between the two %
	pictures, 
indicating intermediate strength of the {\Ox{Sr}--H} ``bonds'', 
which explains the observed stability of the {H-SCO} phase, 
and the reversibility \cite{TriStateXform} of the {H-SCO} and {BM-SCO} %
	interconversion. 
This suggests high mobility of H in SCO, 
which would be invaluable towards energetic and catalytic applications.


\subsection{Carrier-mediated magnetism}
\label{sect:HoleFM}

Finally, we proceed to present a proof-of-concept showing how the measured %
	weak FM in {H-SCO} \cite{TriStateXform} could have been caused by holes. 
As a demonstration, we calculated the FM and AFM cell free energies after the %
	introduction of a hole and an electron, 
holding the lattice vectors constant to better capture the dilute-limit %
	behavior. 

As is evident from Table~\ref{table:CarrierEnergies}, 
FM is favored upon the introduction of holes, 
in contrast to the charge-neutral case; 
whereas AFM is favored when the system is electron-doped, 
as with the neutral case. 
This is consistent with the band coupling-model picture of charge carrier-%
	mediated FM proposed by Dalpian and Wei \cite{CarrierFM}. 
In a real-life scenario, such holes could easily have been introduced by %
	lattice defects and/or dopants (e.g. stray ions), 
which is almost inevitable in the liquid ion-gated environment employed in 
	[\onlinecite{TriStateXform}]. 
Further numerical investigation of point defects in {H-SCO} would be %
	instructive towards the understanding of the observed FM. 
\begin{table}
	\caption{Table of relative cell free energies ${\Delta F}$ under different %
		spin textures and number change of electrons ${\Delta N_{\text{e}}}$. }
	\label{table:CarrierEnergies}
	\begin{ruledtabular}
	\begin{tabular}{ccc}
		${\Delta N_{\text{e}}}$ & $F_{\text{FM}}$ (eV) & $F_{\text{AFM}}$ (eV) %
			\\ \hline
		$0$ (charge-neutral) & ${+0.44}$ & $0$ \\
		${+1}$ (electron-doped) & ${+6.09}$ & ${+5.84}$ \\
		${-1}$ (hole-doped) & ${-3.56}$ & ${-3.38}$
	\end{tabular}
	\end{ruledtabular}
\end{table}

We also note the possibility of asymmetric (or Dzyaloshinsky--Moriya) %
	\cite{AsymmInteraction1,AsymmInteraction2} interaction-mediated effects, 
which often give rise to weak measured FM in otherwise predominantly AFM %
	systems; 
however, evaluation of such interaction would require a full non-collinear %
	treatment of the electron spins, 
which is beyond the scope of this investigation. 


\section{Conclusions}
\label{sect:Conclusions}

By a combination of {DFT+$U$} investigation, ECM, and Hubbard model-based %
	arguments, 
we have analyzed the structural, electronic and magnetic properties of %
	{BM-SCO} and {H-SCO}. 
Such a study is fundamental for and instrumental towards the further %
	refinement, tuning, and application of the SCO system, 
which has already shown promising prospects. 

Moreover, by the treatment of the OVCs in SCO, which are line defects, 
as inner surfaces, 
we have successfully applied ECM to a defect complex in the material bulk. 
This may open up a new perspective towards the understanding of new materials %
	with similar features. 

Particularly, we have shown {H-SCO} to be an interesting material in its own %
	right, 
owing to the moderate bonding strength of the incorporated hydrogen, 
and the increased bandgap energy and conduction electron mobility. 
This may even suggest an entirely new dimension towards the search and %
	development of new classes of materials, 
with incorporated hydrogen as a useful tuning parameter, 
both electronically and magnetically. 


\begin{acknowledgments}
We thank P. Yu for his insightful input throughout the research process. 
We also thank the University Grants Committee of Hong Kong for their Early %
	Career Scheme grant (numbered 24300814), 
\change{the Research Grants Council of Hong Kong for their General Research %
	Fund grant (numbered 14307018), }
and The Chinese University of Hong Kong for their direct grant (numbered %
	4053084), 
all in partial support of this study. 
\end{acknowledgments}


\if\buildbib1
	\bibliography{%
		../../references/defect_ref.bib,%
		../../references/SCO_ref.bib,%
		../../references/general_ref.bib,%
		../../references/vasp_ref.bib}
\else	
%

\fi

\end{document}